\documentclass[a4paper,fleqn,usenatbib,useAMS]{mnras}

\usepackage{graphicx}	% Including figure files
\usepackage{amsmath}	% Advanced maths commands
\usepackage{amssymb}	% Extra maths symbols
\usepackage{multicol}        % Multi-column entries in tables
\usepackage{longtable}
\usepackage{natbib}
\usepackage{wrapfig}
\usepackage{array}

\usepackage{bm}		% Bold maths symbols, including upright Greek
\usepackage{pdflscape}	% Landscape pages

\usepackage{enumerate}
\usepackage[shortlabels]{enumitem}
\usepackage{soul}
\usepackage[dvipsnames]{xcolor}
\usepackage{rotating}
\newcommand{\gsim}{{\;\raise0.3ex\hbox{$>$\kern-0.75em\raise-1.1ex\hbox{$\sim$}}\;}}
\newcommand{\lsim}{{\;\raise0.3ex\hbox{$<$\kern-0.75em\raise-1.1ex\hbox{$\sim$}}\;}}

\defcitealias{Staveley-Smith_1997}{SS97}
\defcitealias{Stanimirovic_1999}{S99}

\usepackage[T1]{fontenc}
\usepackage{ae,aecompl}

\usepackage{newtxtext,newtxmath}
\title[Magnetised superbubbles in the SMC]{Magnetised HI superbubbles in the Small Magellanic Cloud revealed by the POSSUM pilot survey}
\author[Jung et al.]{S. Lyla Jung$^{1}$\thanks{e-mail: \href{mailto:lyla.jung@physics.ox.ac.uk}{lyla.jung@physics.ox.ac.uk}}, A. Seta$^{2}$, J. M. Price$^{2}$, N. M. McClure-Griffiths$^{2}$, J. D. Livingston$^{3}$, B. M. Gaensler$^{4,5,6}$,  \newauthor Y. K. Ma$^{2}$, M. Tahani$^{7}$, C. S. Anderson$^{2}$, C. Federrath$^{2, 8}$, C. L. Van Eck$^{2}$, D. Leahy$^{9}$, S. P. O'Sullivan$^{10}$,\newauthor J. West$^{11}$, G. Heald$^{12}$, T. Akahori$^{13}$
\\
\\
$^{1}$ Department of Physics, University of Oxford, Keble Road, Oxford OX1 3RH, United Kingdom\\
$^{2}$ Research School of Astronomy \& Astrophysics, The Australian National University, Canberra ACT 2611, Australia
\\
$^{3}$ Max Planck Institute for Radio Astronomy, Auf dem H\"ugel 69, Bonn D-53121, Germany\\
$^{4}$ Department of Astronomy and Astrophysics, University of California Santa Cruz, 1156 High Street, Santa Cruz, CA 95060, USA\\
$^{5}$ Dunlap Institute for Astronomy and Astrophysics, University of Toronto, 50 St. George Street, Toronto, ON M5S 3H4, Canada \\
$^{6}$ David A. Dunlap Department of Astronomy and Astrophysics, University of Toronto, 50 St. George Street, Toronto, ON M5S 3H4, Canada \\
$^{7}$ Banting and KIPAC Fellowships: Kavli Institute for Particle Astrophysics \& Cosmology (KIPAC), Stanford University, Stanford, CA 94305, USA\\
$^{8}$ Australian Research Council Centre of Excellence in All Sky Astrophysics (ASTRO3D), Canberra, ACT 2611, Australia\\
$^{9}$ Department of Physics and Astronomy, University of Calgary, Calgary T2N 1N4, Canada\\
$^{10}$ Departamento de Física de la Tierra y Astrofísica \& IPARCOS-UCM, Universidad Complutense de Madrid, 28040 Madrid, Spain\\
$^{11}$ Dominion Radio Astrophysical Observatory, Herzberg Astronomy and Astrophysics, National Research Council Canada, PO Box 248, Penticton, BC V2A 6J9, Canada\\
$^{12}$ Australia Telescope National Facility, CSIRO, Space and Astronomy, PO Box 1130, Bentley WA 6102, Australia\\
$^{13}$ Mizusawa VLBI Observatory, National Astronomical Observatory of Japan, 2-21-1, Osawa, Mitaka, Tokyo 181-8588, Japan
}

\begin{document}
\maketitle

\begin{abstract}
Neutral hydrogen (HI) bubbles and shells are common in the interstellar medium (ISM). Studying their properties provides insight into the characteristics of the local ISM as well as the galaxy in which the bubbles reside. We report the detection of magnetic fields associated with superbubbles in the nearby irregular galaxy, the Small Magellanic Cloud (SMC). Using the Polarisation Sky Survey of the Universe's Magnetism (POSSUM) pilot survey, we obtain a high-density grid ($\approx 25 \,\rm sources\,deg^{-2}$) of Faraday rotation measure (RM) from polarized sources behind the SMC. This provides a sufficiently large number of RM measurements to study the magnetic properties of three of the largest HI shells previously identified in the SMC. The RM profiles as a function of distance from the shell centre show characteristic patterns at angular scales comparable to the shell size. We demonstrate that this can be explained by magneto-hydrodynamic simulation models of bubbles expanding in magnetised environments. From the observations, we estimate the line-of-sight magnetic field strength at the edges of the shells is enhanced by $\sim1\,\rm \mu G$ with respect to their centres. This is an order of magnitude larger than the field strength in the ambient medium ($\sim 0.1\,\rm \mu G$) estimated based on the expansion velocity of the shells. This paper highlights the power of densely mapped RM grids in studying the magnetic properties of galactic substructures beyond the Milky Way.
\end{abstract}

\section{Introduction}

High-resolution neutral hydrogen (HI) observations demonstrate that structures with shell-like morphology are ubiquitous in the interstellar medium (ISM) of gas-rich galaxies. 
These HI shells appear in a wide range of sizes from hundreds of parsecs to several kiloparsecs
(\citealt{Heiles_1979}, \citealt[hereafter SS97]{Staveley-Smith_1997}, \citealt{Bagetakos_2011}).
The size of the shells changes with the Doppler shift, indicating that they are expanding into the surrounding ISM. The expansion is powered by sequential bursts of stellar winds and supernovae from OB stellar associations. Studying the properties of superbubbles provides valuable information about the stellar feedback mechanism as well as how the local ISM is connected to the large-scale galactic environment (\citealt{McClure-Griffiths_2002}). 

Earlier theoretical studies and numerical simulations show that magnetic fields shape the overall morphology of bubbles while stabilising the surface from fragmentation due to the growth of hydrodynamic instabilities (e.g., \citealt{Ferriere_1991, Tomisaka_1998, Hanayama_2006, Stil_2009, Kim_2015, Ntormousi_2017, Gentry_2019}). The expansion of a bubble results in the compression of the field lines in the surrounding medium, which enhances the magnetic field strength locally in the medium. This, depending on the energy distributions and ambient magnetic field properties, may result in magnetic fields tangential to the boundary of the bubble. Therefore, the magnetic field component perpendicular to the direction of expansion becomes stronger. The magnetic pressure hinders expansion in the direction perpendicular to the magnetic fields, especially when a strong, large-scale magnetic field exists in the surrounding ISM (\citealt{Chen_2022}). 

Faraday rotation of polarized background emission is an observational tracer of magnetic fields in the ionised gas. The degree of Faraday rotation is proportional to RM$\lambda^{2}$, where $\lambda$ is the observing wavelength, and RM is the rotation measure described as follows: 
\begin{equation}\label{eq:RM} 
    \left(\frac{\rm RM}{\rm rad\,m^{-2}}\right) = \mathcal{C}\int_{\rm observer}^{\rm source} \left(\frac{n_{\rm e}}{\rm cm^{-3}}\right)\left(\frac{\rm B_{\parallel}}{\mu G}\right)\left(\frac{{\rm d}r}{\rm pc}\right),
\end{equation}
where $\mathcal{C} = 0.812\,\rm rad\,m^{-2}pc^{-1}cm^{3}\mu G^{-1}$, $n_{\rm e}$ is electron density, $B_{\parallel}$ is the magnetic field strength parallel to the line-of-sight, and $r$ is a path length along the sightline. 
The compression at expanding shells provides high density and magnetic field strength, naturally leading to enhanced Faraday rotation in the region. 

\citet{Stil_2009} predict observable patterns of magnetised bubbles in the RM grid of polarized sources using magneto-hydrodynamic simulations. According to their study, the pattern in which a 3D magnetic field around a bubble imprints on the observable RM distribution varies significantly depending on the viewing angle. When the global magnetic fields are perpendicular to the sightline, the Faraday rotation at the far and near sides of a spherically symmetric bubble nearly cancel out, and the observable RM is close to zero. When the line-of-sight magnetic field is dominant over the plane-of-sky component, the estimated RM is overall enhanced along the bubble's boundary with respect to the surrounding area. Note that their models only assume coherent uniform magnetic fields in the medium surrounding the bubbles. We will show in later sections that the observable RM patterns around bubbles can also change when random magnetic fields are introduced.

There are observational studies investigating the magneto-ionic properties of ISM shells and the surrounding gas environment (e.g., \citealt{Vallee_1983, West_2007, Whiting_2009, Harvey-Smith_2011, Heald_2012, Gao_2015, Purcell_2015, Mulcahy_2017, Costa_2018, Thomson_2018, Thomson_2019}). Thus far, the polarized source densities of existing point-source RM catalogues ($\sim 1\,\rm deg^{-2}$) have been only sufficient to sample bubbles that are nearby (i.e., Milky Way objects). At a given RM grid source density, objects closer to the observer and more extended on the sky are generally sampled with higher statistical significance as examined in detail by \citet{Jung_2023b}.

In this paper, we study the magnetic field amplification process via bubble expansion using radio polarization observations for shells in the nearby irregular galaxy, the Small Magellanic Cloud (SMC). 
Being one of the nearest gas-rich neighbouring galaxies of the Milky Way, the SMC provides an opportunity to study resolved gaseous structures in the ISM of an external galaxy. The ISM of the SMC is inhomogeneous, filled with HI shells and bubbles that have been catalogued extensively (\citetalias{Staveley-Smith_1997}, \citealt[hereafter S99]{Stanimirovic_1999}, \citealt{Hatzidimitriou_2005}). We fully utilise the new high-source-density RM grid of the Polarisation Sky Survey of the Universe’s Magnetism (POSSUM) using the Australian Square Kilometre Array Pathfinder (ASKAP) to search for magnetised superbubbles in the SMC. The angular sizes of some of the largest catalogued supergiant shells (SGS) are larger than a degree; therefore, POSSUM ensures that each of these shells overlaps with a significant number of polarized extragalactic point sources.

The paper is structured as follows. We present an overview of the data used in this study in Section \ref{sec:method}. Section \ref{sec:results} presents our inspections of the RM distribution around three of the largest HI superbubbles in the SMC. In Section \ref{sec:model}, we explore the connection between the global magnetic field and the pattern appearing in the RM grid using numerical simulations. The magnetic field strengths inferred from the observations are presented in Section \ref{sec:B_strength}. Section \ref{sec:summary} is the summary.

\section{Method}\label{sec:method}

This study draws information from polarised radio sources behind the SMC and H$\alpha$ and HI emission from the SMC's ISM to study magnetic fields associated with HI shells in the SMC. In Section \ref{sec:method-rm}, we give an overview of the POSSUM pilot RM catalogue and explain our method for correcting the Faraday rotation caused by the Galactic foreground. Section \ref{sec:method-halpha} describes the H$\alpha$ data we use to estimate the extent of diffuse ionised ISM of the SMC. In Section \ref{sec:method-HI}, we introduce the HI shell sample and describe the HI data that we use to define the extent of the shells.

\subsection{Faraday rotation of polarised radio sources}\label{sec:method-rm}
\subsubsection{POSSUM pilot observations}\label{sec:method-possum}

POSSUM (\citealt{Gaensler_2010}) is one of the major ongoing surveys of ASKAP (\citealt{Hotan_2021}). 
This study uses the POSSUM pilot data towards the SMC (observation sbid: SB43237) observed in full polarization over $800-1088\,\rm MHz$ with $1\,\rm MHz$ frequency resolution.
ASKAP's uniquely large field-of-view ($\sim 30\,\rm deg^{2}$) covers the SMC and its immediate surroundings with a single pointing centred on $\rm (RA\,J2000, Dec\,J2000) = (1^{h} 01^{m} 18^{s}.666, -72^{\circ} 33' 06''.997)$. The observations were performed on 14 September 2022 for 10 hours of integration time.

The SMC is a radio-loud object with substantial continuum emission detectable by ASKAP. To prevent the obscuration of background sources by the SMC, we subtract the diffuse Stokes I emission from the Stokes I image with a median filter method (see \citealt{Vanderwoude_2024} Section 3.1.1), and subsequently conducted source finding on the diffuse emission subtracted Stokes I image with pyBDSF (\citealt{2015ascl.soft02007M}). Meanwhile, Stokes I, Q, and U cubes are convolved to a common beam (20 arcsec; the beam size at the lowest frequency) over the channels and corrected for ionospheric Faraday rotation. It was then processed using the POSSUM pipeline (\citealt{Gaensler_inprep} in preparation; \citealt{VanEck_inprep} in preparation) to measure the polarization properties of the sources; we summarize the pipeline's steps as the following. For each Stokes I source within the catalog produced by pyBDSF, a cutout around the source location is extracted and the polarized diffuse emission is subtracted on a per-Stokes and per-channel basis as median diffuse polarised emission within an annulus around the source, with the inner radius set to 17 arcsec and outer radius to 109 arcsec \citep{oberhelman24}. The spectra are then processed with 1D RM synthesis (\citealt{Brentjens_2005}) using the {\sc RM Tools 1D} software (\citealt{Purcell_2020}), which outputs values for the polarization properties of the source. The results from all sources are collected into a single catalog.
To prevent duplicated detection of sources, sources located closer to each other than 36 arcsec are removed from the catalogue, based on their signal-to-noise ratio in polarized intensity (\citealt{Price_inprep} in preparation).

For this paper, we use a total of 1006 sources with a polarized signal-to-noise ratio above seven and a polarization fraction between 0.01 and 0.5. 
Typical errors of the sources in the Stokes I and polarization intensity are $\approx 38\,\rm \mu Jy\,beam^{-1}$ and $\approx 21\,\rm \mu Jy\,beam^{-1}$, respectively.
The high sensitivity of ASKAP enables the average polarized source density of $\approx 25\,\rm deg^{-2}$ in the field. This is the highest-density RM grid towards the SMC thus far, achieving an order of magnitude increase compared to the previous RM grid studies of the region \citep[$\sim 1\,\rm deg^{-2}$ in][]{Mao_2008, Livingston_2022}. 
Compared to the full band 1 POSSUM survey results (\citealt{Vanderwoude_2024}), the polarized source density toward the SMC is lower than the typical source density in the extragalactic regions
($\approx 35\,\rm deg^{-2}$) and higher than that of the Galactic plane region ($\approx 13.5\,\rm deg^{-2}$). This is in part because the SMC is depolarizing the background sources. Further investigations on this topic will be followed in \citealt{Price_inprep} (in preparation) where the authors perform a statistical analysis of the polarized source density across the SMC in conjunction with Stokes Q and U model fitting and RM structure functions, ascribing the depolarization to a turbulent magneto-ionic medium along the line-of-sight.

\begin{figure}
    \centering
    \includegraphics[width=\columnwidth]{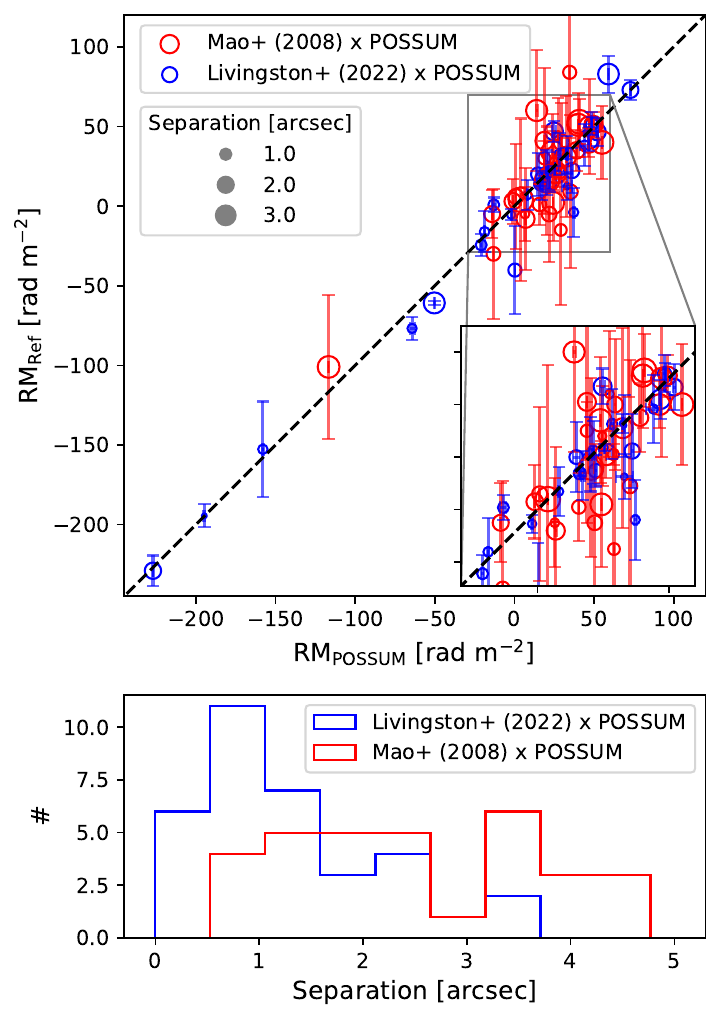}
    \caption{
    The cross-match between RM measurements in this study and previous RM catalogues (red: \citealt{Mao_2008} and blue: \citealt{Livingston_2022}). The upper panel compares the RM values. The dashed line shows the one-to-one relation for reference, and the area of the circles is proportional to the separation between the matched pairs. The error bars show the RM measurement error. A zoom-in panel is provided for better visualisation at RM ranges where the circles are highly clustered. 
    The bottom panel shows the distribution of angular separation between the matched pairs. 
    Overall, the RMs of cross-matched sources agree with the previous measurements. 
    }  
    \label{fig:matching}
\end{figure}

We cross-match the POSSUM sources to existing RM catalogues and compare the RM values (see the upper panel of Fig. \ref{fig:matching}). In total, 32 and 34 sources from the POSSUM pilot observations are matched with \citet[red circles]{Mao_2008} and \citet[blue circles]{Livingston_2022} catalogues, respectively, with the maximum separation of the match set to 5 arcsec. The error bars show the RM measurement error from each catalogue. Note the errors from POSSUM observations (i.e., error bars along the x-axis) are much smaller than the axis scale of the figure and, therefore, hardly visible. The median uncertainties of the sources from \citet{Mao_2008}, \citet{Livingston_2022}, and the POSSUM data are $24.5$, $7.65$, and $1.65\,\rm rad\,m^{-2}$. All the cross-matched POSSUM sources have nearly identical RM values in comparison to previous studies, especially considering the large RM error in the previous measurements. The diameter of the circles is proportional to the angular separation between the matched pairs. The histograms of the separations are presented in the bottom panel of Fig. \ref{fig:matching}.

\subsubsection{Galactic foreground RM model}\label{sec:foreground}

The observed RM is a superposition of any Faraday rotation occurring between the source and the observer (equation \ref{eq:RM}). 
At angular scales of interest in this study, much of the spatial variation in the RM distribution comes from the Milky Way's magnetised ISM. 
The contribution intrinsic to the polarized sources and that of the intergalactic medium is at much smaller scales in the dispersion of $\sim 6-8\,\rm rad\,m^{-2}$ (\citealt{Schnitzeler_2010, Han_2017, Taylor_2024}).

We construct the Milky Way RM model following the method introduced by \citet{Mao_2008} to correct for the Milky Way's foreground in the region towards the SMC\footnote{Large-scale Milky Way RM models (e.g., \citealt{Hutschenreuter_2022}) are not suitable to remove the foreground at the relatively small angular scale of interest in this study.}. Using the POSSUM sources that are close to but outside most emission from the SMC ($N_{\rm HI, SMC} < 2\times10^{21}\,\rm cm^{-2}$ and $I_{\rm SMC, H\alpha}<2.5\,\rm rayleighs$, where $N_{\rm HI, SMC}$ is the HI column density and $I_{\rm SMC, H\alpha}$ is the H$\alpha$ intensity of the SMC), we perform a least-square fit to the observed RM values as a function of the right ascension in degrees: 
\begin{equation}
    \left(\frac{\rm RM_{MW}}{\rm rad\,m^{-2}}\right) = 29.18-3.60\times{\rm RA}\cos({\rm Dec}),
\end{equation} 
where $\rm RA$ and $\rm Dec$ are the J2000 right ascension and declination in degrees.

Note that the actual distribution of the diffuse ionised corona of the SMC can reach far beyond its main emission region, out to $\lsim 35\,\rm kpc$ from the centre of the Magellanic System (\citealt{Smart_2019, Krishnarao_2022, Lucchini_2024}). It is likely that the Milky Way foreground model of RM incorporates the contributions of such extended structures of the SMC to some degree. Constraining the true extent of the SMC's magnetised corona is beyond the scope of this paper and will be covered in a follow-up study using the full POSSUM survey.

\subsection{Diffuse H$\alpha$ emission}\label{sec:method-halpha}
 
\begin{figure}
    \centering
    \includegraphics[width=\columnwidth]{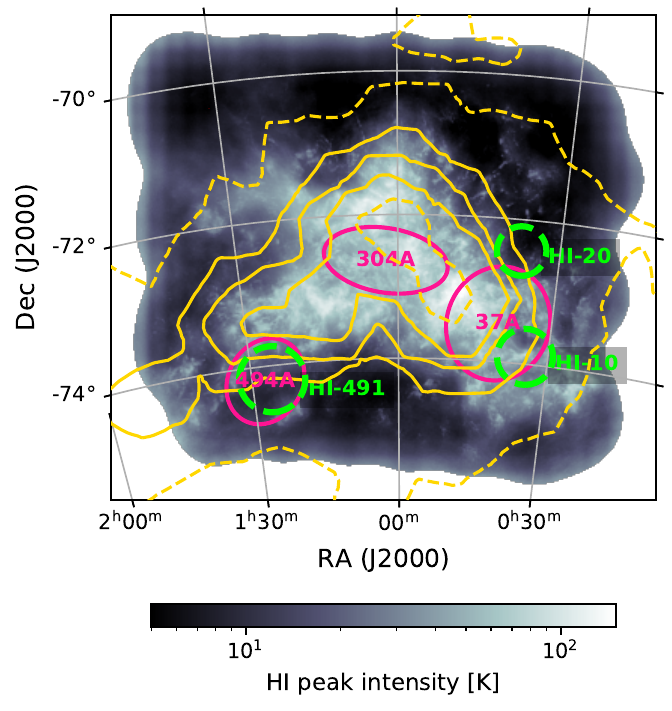}
    \caption{
    The distribution of HI shells in the SMC overlaid on the top of the HI peak intensity distribution (grey scale; 30 arcsec synthesized beam). The HI shells selected for this study are shown as green dashed-line circles. Pink ellipses are SGS in the \citetalias{Stanimirovic_1999} catalogue. 
    The yellow contour lines show $I_{\rm H\alpha, SMC}=$ 0.3 and 30 rayleighs (dashed lines) and the levels in between increasing by 0.5 dexes in a log scale (solid lines). 
    }  
    \label{fig:shell_sample}
\end{figure}

\citet{Smart_2019} provide a map of the extended H$\alpha$ emission of the SMC observed with the Wisconsin H$\alpha$ Mapper (WHAM). Their results reveal the diffuse ionised gas in the gaseous halo surrounding the SMC. We show the distribution of the H$\alpha$ intensity ($I_{\rm SMC, H\alpha}$) in Fig. \ref{fig:shell_sample} with yellow contour lines increasing from 0.3 rayleighs (outer dashed line) to 30 rayleighs (inner dashed line) in log-scale. The WHAM observations trace diffuse large-scale H$\alpha$ emission, notably from the SMC Bar region and the Wing region\footnote{See Fig. 1 of \citet{McClure-Griffiths_2018} for a visual reference for the SMC Bar and Wing regions.}. Potential substructures in H$\alpha$ emission, such as the HI shells, are smoothed out due to the large angular resolution. 

In addition to the HI shells of interest, the SMC contains a wealth of structures that produce strong Faraday rotation. 
We use the H$\alpha$ intensity as a proxy for the extent of the SMC's ionised ISM that may produce strong Faraday rotation unrelated to the HI shells of interest. We define an area at the outskirts of the SMC with low H$\alpha$ intensity (arbitrarily chosen criterion of $I_{\rm SMC, H\alpha}<2\,\rm rayleighs$) and consider RM measurements in this region are relatively free from Faraday rotation within interlopers. See Section \ref{sec:results} for further discussion of the impact of this threshold on the results presented in this paper.

\citet{Smart_2019} discuss how the H$\alpha$ emission from the SMC undergoes extinction by dust in the Milky Way as well as the SMC itself. They predict that the true H$\alpha$ intensity in the region surrounding the SMC's main body is $12.9-14.4\%$ ($20-23\%$) higher before the Milky Way foreground (SMC internal) extinction. 
Since we are only interested in the outer regions of the SMC with low gas and dust column densities, we expect the effect of extinction by the SMC itself to be low. For simplicity, we continue using the uncorrected $I_{\rm SMC, H\alpha}$ because the exact value of $I_{\rm SMC, H\alpha}$ after the Milky Way extinction correction does not affect the results presented in this paper.

\subsection{HI emission}\label{sec:method-HI}
\subsubsection{GASKAP survey}
The Galactic ASKAP survey (GASKAP; \citealt{Dickey_2013, Pingel_2022}) maps the distribution of HI and OH in the Milky Way and the Magellanic System. The GASKAP-HI SMC data has $1.1\,\rm K$ per channel sensitivity, 30 arcsec synthesized beam, and $0.98\,\rm km\,s^{-1}$ velocity resolution (\citealt{Pingel_2022, McClure-Griffiths_2018, Ma_2023}). In Fig. \ref{fig:shell_sample}, we show the HI peak intensity map from the GASKAP-HI (grey scale). 
In Section \ref{sec:results}, we use the GASKAP-HI data cube to obtain an up-to-date view of the morphological and kinematic structures of selected HI shells. Furthermore, in Section \ref{sec:B_strength}, we use the HI column density distribution to estimate the magnetic field strength from the RM measurements. 

\subsubsection{HI shell sample}\label{sec:shell_catalogue}

\begin{figure*}
    \centering
    \includegraphics[width=\textwidth]{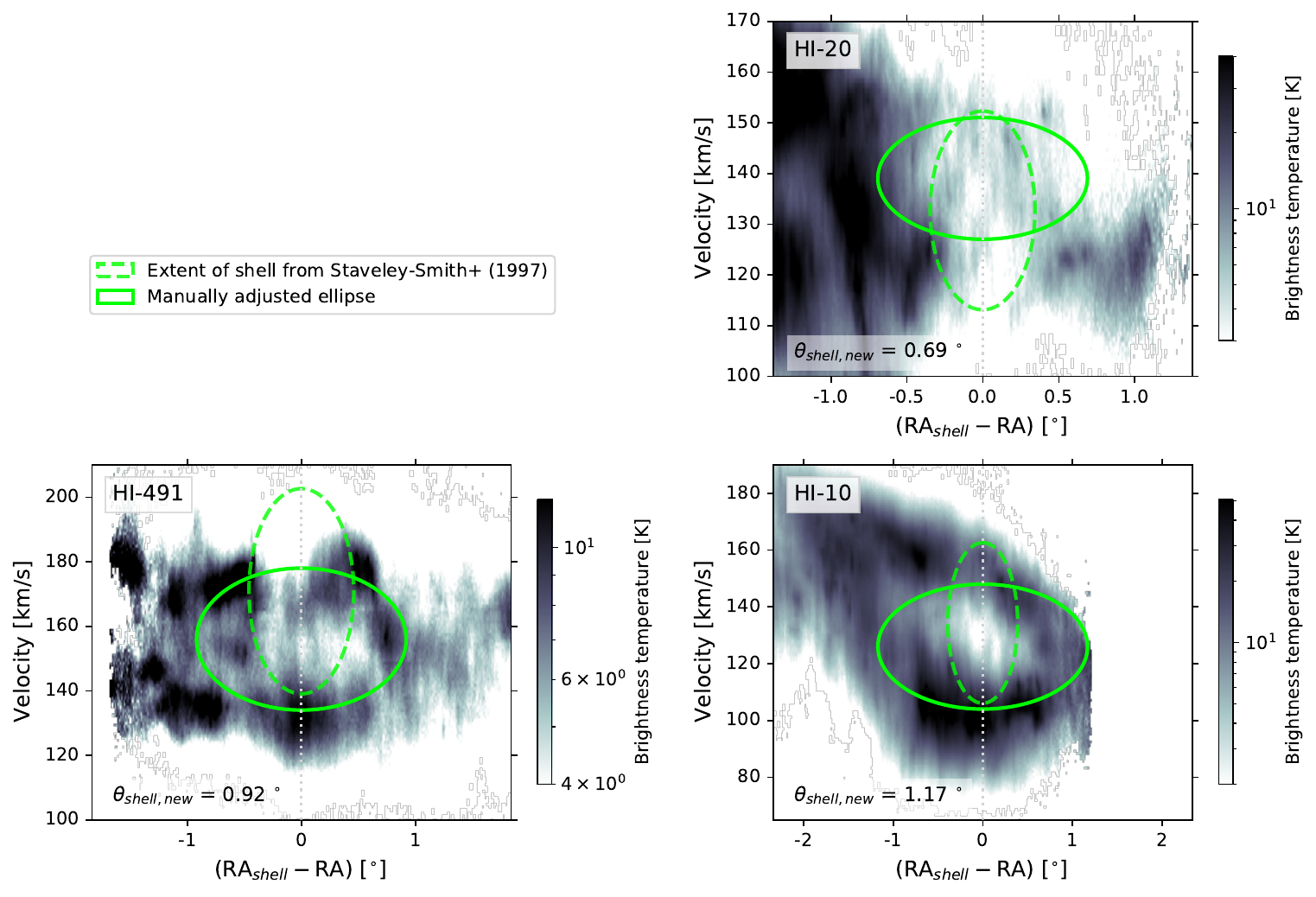}
    \caption{The position--velocity distribution of HI shells from the GASKAP-HI SMC cube. We show the HI emission within a long thin slit, $[4\times0.1]\,\theta_{\rm shell, new}$, that passes the centre of the HI shell and whose long side is aligned along the RA (J2000). The dashed ellipse shows the angular size and the expansion velocity identified by \citetalias{Staveley-Smith_1997}. The grey dotted vertical line shows the zero offset along the RA (J2000). We redefine the full extent of the HI shells by manually adjusting the ellipse in the position--velocity space while the central RA of the shell is fixed (ellipse with the solid-line boundary).
    }  
    \label{fig:pv}
\end{figure*}

There are several previous studies cataloguing shell-like structures in the SMC. \citetalias{Staveley-Smith_1997} identify 501 HI shells in total: most of them (343/501) were visually identified based on the expanding kinematic characteristics in the RA (J2000) -- velocity space.

We focus on three of the largest HI-shells for further analysis: HI-10, HI-20, and HI-491. They are highlighted as green dashed line ellipses in Fig. \ref{fig:shell_sample}. The selected shells are located on the outskirts of the SMC, away from the main SMC bar region. Restricting the analysis to these HI shells with large angular sizes at the outskirts of the SMC ensures a sufficient number of RM measurements that are relatively free from overlapping objects around individual bubbles and provides a good statistical significance for the RM grid analysis.

The HI-10 and HI-491 shells are later identified by \citetalias{Stanimirovic_1999} to be substructures of larger HI supergiant shells (sgs-37A and sgs-494A, respectively, pink ellipses). 
%We stick to using the shell properties (e.g., position and radius) identified by \citetalias{Staveley-Smith_1997} for simplicity, yet our analysis extends to the full spatial extent of the host SGS.
This is because the data used by \citetalias{Staveley-Smith_1997} to define the shells were not sensitive to large structures.

In this regard, we redefine the angular extent of the shells ($\theta_{\rm shell, new}$) using the GASKAP-HI data while keeping the centre of the shells fixed to the \citetalias{Staveley-Smith_1997} definitions. Fig. \ref{fig:pv} shows the position--velocity diagram of HI emission along a slit ($4\theta_{\rm shell, new}$ length and $0.1\theta_{\rm shell, new}$ width) that passes through the centre of each HI shell and whose long side is aligned along the RA (J2000)\footnote{Although the extents of the shells are determined within slits aligned along the RA axis, we confirm that the by-eye measurements are valid for position-velocity slices along other axes too, including the axis along which the RM profiles are measured in Section \ref{sec:results}.} (HI-10: bottom right, HI-20: top right, and HI-491: bottom left). The dashed line ellipse is the extent of the shell identified by \citetalias{Staveley-Smith_1997}, and the solid line ellipse is the newly defined shell by manually adjusting the ellipse to match the HI emission better. For reference, $\theta_{\rm shell, new}=3\times\theta_{\rm shell, SS97}$ for HI-10, $\theta_{\rm shell, new}=2\times\theta_{\rm shell, SS97}$ for HI-20 and HI-491. The newly identified expansion velocities of HI-10, 20, and 491 shells, i.e., the extent of the solid line ellipse along the y-axis, are $22, 12$, and $22\,\rm km\,s^{-1}$, respectively.

\section{Result}\label{sec:results}

\begin{table*}
\caption{Summary of the observed properties and the magnetic field configuration of the HI shells.
Row 1: the SGS associated with the HI shell according to \citetalias{Stanimirovic_1999}. Row 2 and 3: the angular radius according to \citetalias{Staveley-Smith_1997} and the corresponding physical radius assuming a distance of $60\,\rm kpc$. Row 4 and 5: the angular radius redefined for this study (see Fig. \ref{fig:pv}) and the corresponding physical radius.
Row 6: the expansion velocity according to \citetalias{Staveley-Smith_1997}. Row 7: the expansion velocity redefined for this study. 
Row 8: the position angle of the axis of the maximum RM gradient measured in the image plane. Row 9: the difference between the maximum and minimum of the RM profile. Row 10: the reference RM to correct for the Faraday rotation within the SMC away from the bubble (see Section \ref{sec:B_strength} for the justification). Row 11: the maximum line-of-sight magnetic field strength associated with the bubble. 
Row 12: our interpretation of the magnetic field configuration around the bubble based on the simple models presented in Section \ref{sec:model}.
}\label{tab:t1}
\begin{minipage}{\textwidth}\centering
\begin{tabular}{m{3.4cm}m{3.9cm}m{3.9cm}m{3.9cm}}
\hline
 & HI-10 & HI-20 & HI-491 \\ [0.5ex] \hline\hline
 Associated SGS &   sgs-37A    &   -    &    sgs-494A    \\ [0.5ex] \hline
 $\theta_{\rm shell, SS97}$ [arcmin] &   23.4   &   20.7    &    27.6    \\ [0.5ex] \hline
 $r_{\rm shell, SS97}$ [pc] &   408   &   361    &    482    \\ [0.5ex] \hline
 $\theta_{\rm shell, new}$ [arcmin] &   70.2   &   41.4    &    55.2   \\ [0.5ex] \hline
  $r_{\rm shell, new}$ [pc] &   1225   &   723    &    963    \\ [0.5ex] \hline
$v_{\rm exp, SS97}$ [$\rm km\,s^{-1}$] & 28.3  & 19.6 &  31.9   \\ [0.5ex] \hline
 $v_{\rm exp, new}$ [$\rm km\,s^{-1}$] & 22  & 12 & 22  \\ [0.5ex] \hline
 RM gradient position angle [$^{\circ}$]&   31.3    &   113.6    &    125.2    \\ [0.5ex] \hline
 %RM profile shape &   sinusoidal    &    symmetric (U-shaped centre)  &    symmetric (U-shaped centre)   \\ [0.5ex] \hline
 $\max(\Delta \rm RM)$ [$\rm rad\,m^{-2}$] &   $47.3\pm19.4$    &   $30.3\pm18.8$    &    $28.9\pm15.9$    \\ [0.5ex] \hline
 $\rm RM_{ref}$ [$\rm rad\,m^{-2}$]&   -10.2    &    -13.0   &    -21.1    \\[0.5ex]  \hline
 $\max(|B_{\rm \parallel}|)$ [$\mu G$]&   $0.46^{+0.89}_{-0.28}$    &   $1.23^{+0.88}_{-0.56}$    &     $0.63^{+0.28}_{-0.15}$     \\[0.5ex]  \hline
  Magnetic field configuration &
\begin{itemize}[leftmargin=0.5em]
\setlength\itemsep{1em}
\setlength\itemindent{0em} 
    \item Coherent plane-of-sky mean field at scales comparable to or larger than the bubble
    \item The far and near sides are not in perfect symmetry
    \item Substantial random component 
    \vspace{-1em}
\end{itemize}
  &    
\begin{itemize}[leftmargin=0.5em]
\setlength\itemsep{1em}
\setlength\itemindent{0em} 
    \item Substantial line-of-sight coherent mean field at scales comparable to or larger than the bubble
    \item[] and/or 
    \item The far and near sides are largely symmetric
    \item Substantial random component 
    \vspace{-1em}
\end{itemize}
  &     
\begin{itemize}[leftmargin=0.5em]
\setlength\itemsep{1em}
\setlength\itemindent{0em} 
    \item Substantial line-of-sight coherent mean field at scales comparable to or larger than the bubble
    \item The far and near sides may not be in a perfect symmetric
    \item Substantial random component 
    \vspace{-1em}
\end{itemize}
  \\ \hline
\end{tabular}
\end{minipage}
\end{table*}

\begin{figure*}
    \centering
    \begin{turn}{90}\includegraphics[width=1.2\textwidth]{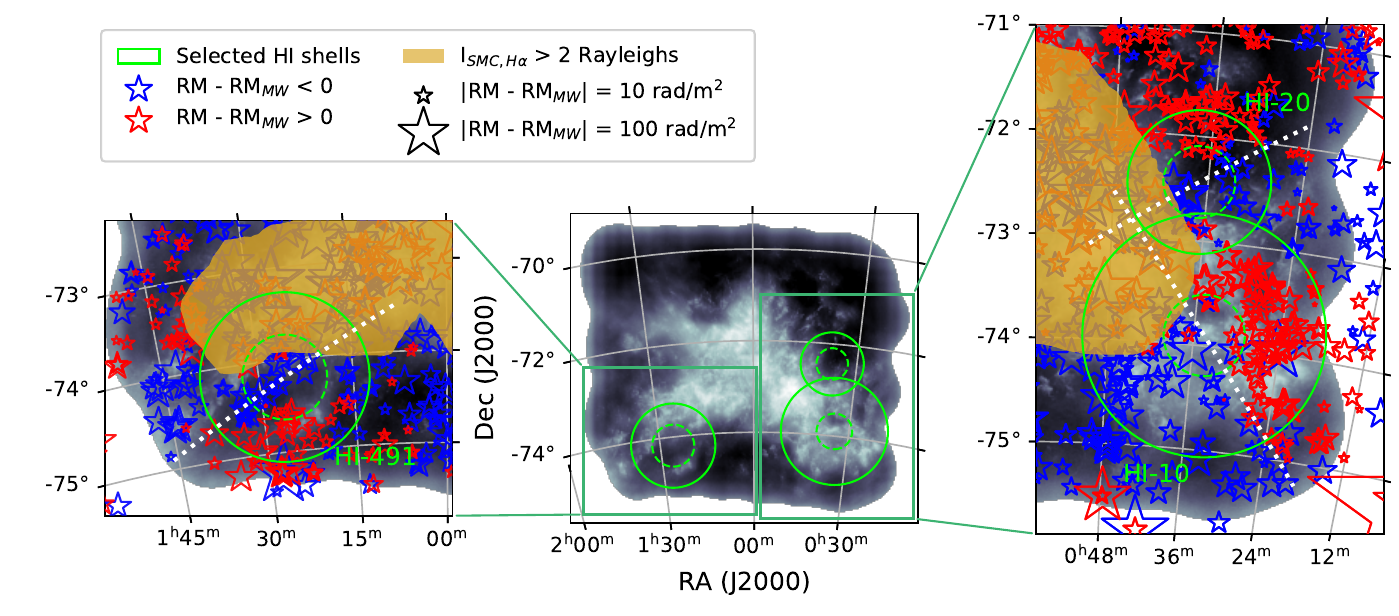}
    \end{turn}
    \caption{
    The foreground corrected RM grid around the selected HI shells. The central panel is the HI peak intensity map of the SMC, and the green circles show the location and the size of HI shells (dashed lines: $\theta_{\rm shell, SS97}$ and solid lines: $\theta_{\rm shell, new}$). The green boxes in the central plot show the locations of the zoom-in panels. The HI-10 and HI-20 shells are presented in one zoom-in panel due to their proximity. On top of the peak HI intensity map in each panel, the star symbols show the location of polarized background sources. The colour of the symbols shows the sign of foreground corrected RM measurements (blue: negative and red: positive) and the size is proportional to the magnitude of RM. The white dotted line crossing each HI shell is the axis of the maximum RM gradient as defined in the text. The yellow shade shows the area with $I_{\rm SMC, H\alpha}>2\,\rm rayleighs$.}  
    \label{fig:bubbles_zoom}
\end{figure*}

In this section, we examine patterns in the RM grid around the selected HI bubbles: HI-10, HI-20, and HI-491. The basic properties of the shells and our findings presented in this and the following sections are summarised in Table \ref{tab:t1}. 

Fig. \ref{fig:bubbles_zoom} shows the three HI shells on top of the HI peak intensity distribution (grey-scale background). Each HI shell corresponds to two circles in green colour. The circle with a dashed boundary shows the angular extent of the shells identified by \citetalias{Staveley-Smith_1997}, and the circle with a solid boundary shows the newly defined extent described in Section \ref{sec:shell_catalogue}. The RM measurements are presented as star symbols in each panel. The colour and the area of the symbols correspond to the sign (red: positive, blue: negative) and the absolute value of the foreground subtracted RM ($=\rm RM - RM_{\rm MW}$).
The yellow shaded region marks the area where $I_{\rm SMC, H\alpha}$ is larger than 2 rayleighs. 

For each HI shell, we identify the axis that maximizes the RM gradient (white dotted lines in Fig. \ref{fig:bubbles_zoom}). This is done by taking RM values within a certain sized aperture centred at the HI shell and identifying the position angle of the axis that maximises the RM error-weighted average RM values on each side of the axis. As mentioned in Section \ref{sec:method-halpha}, we exclude any RM measurements in the strong H$\alpha$ emission region ($I_{\rm SMC, H\alpha}>2\,\rm rayleighs$) from this process. We obtain a set of position angles by varying the angular size of the aperture between $[0.5-1.5]\, \theta_{\rm shell, new}$ with an increment of $0.1\, \theta_{\rm shell, new}$. Although the position angle fluctuates depending on the aperture size, we take the median value as the representative position angle of the RM gradient axis for further analysis. 

\begin{figure*}
    \centering
    \includegraphics[width=0.85\textwidth]{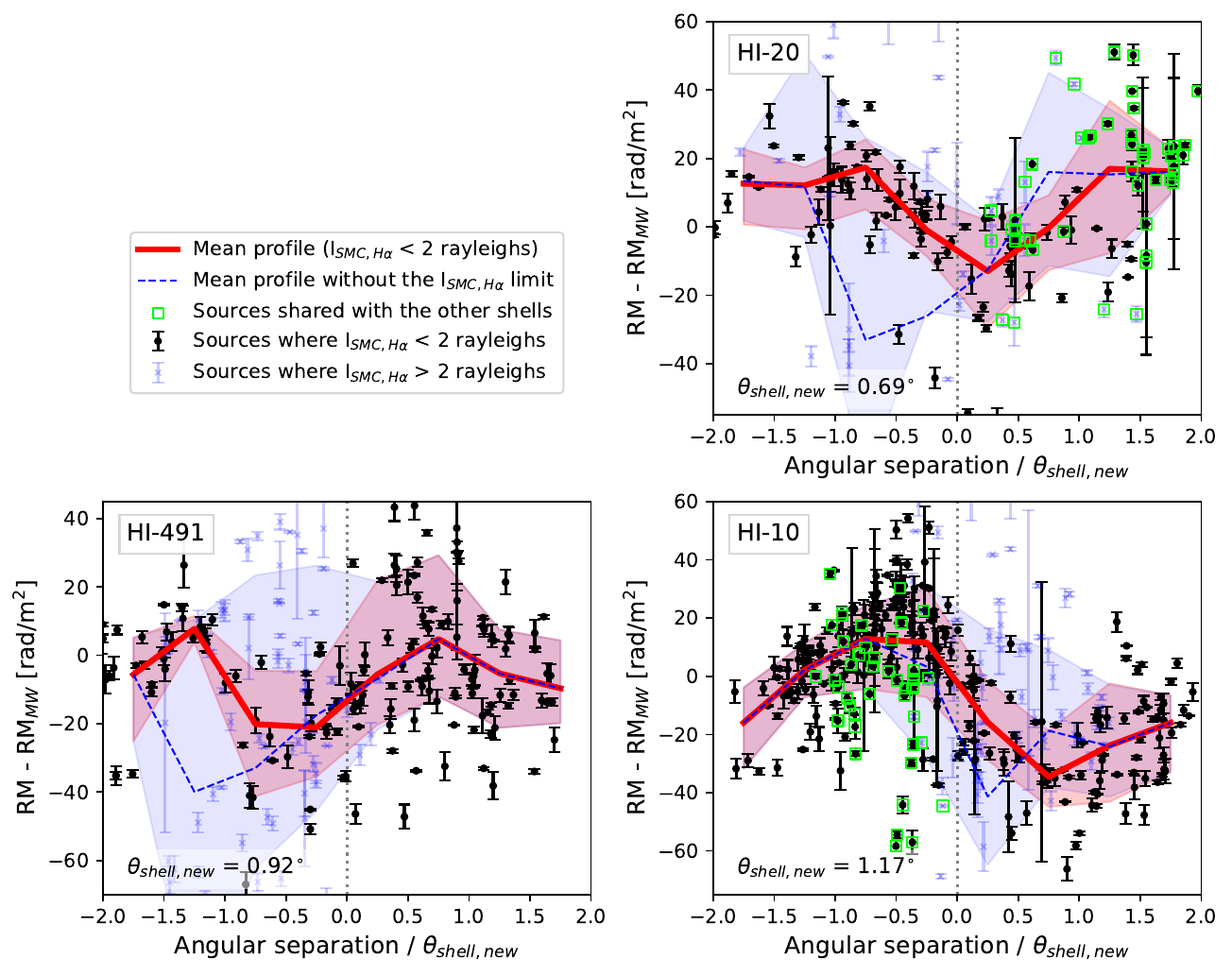}
    \caption{
    Foreground corrected RM profiles across HI shells (HI-10: bottom right, HI-20: top right, and HI-491: bottom left). The x-axis is the angular separation from the shell centre along the direction perpendicular to the RM gradient axis (the white dotted line shown in Fig. \ref{fig:bubbles_zoom}) normalised by $\theta_{\rm shell, new}$ of each shell. 
    The black points with error bars are the RM measurements in the low H$\alpha$ emission region ($I_{\rm SMC, H\alpha}<2\,\rm rayleighs$). We use these points to obtain the mean RM profile and its $1\sigma$ scatter (red line and shade) binned in $0.5\,\theta_{\rm shell, new}$ width between $[-2, 2]\,\theta_{\rm shell, new}$ range. The blue points are the RM sources overlapping with the strong H$\alpha$ emission region. RM profiles obtained without limiting the H$\alpha$ intensity (blue dashed line and shade) fluctuate with a large scatter. The green squares in the HI-10 and HI-20 panels show the RM sources shared with the other shell. }  
    \label{fig:main}
\end{figure*}

We present our visual inspection of the RM distribution around the bubbles as well as the RM profile as a function of the angular offset from the axis of the maximum RM gradient (see Fig. \ref{fig:main}). The motivation to do so is that an expanding bubble would create a strong gradient in the RM distribution at angular scales similar to its size (e.g., \citealt{Heald_2012}).
In Section \ref{sec:model}, we use simple numerical models of expanding bubbles to show that this is indeed true for specific configurations of ambient magnetic field. 
We have confirmed that the overall shapes of the RM profiles presented below are independent of the Milky Way foreground RM subtraction process described in Section \ref{sec:foreground}.

\subsection{HI-10 shell}

There are 374 RM measurements in total within $2\,\theta_{\rm shell, new}$ of the HI-10 shell centre (328 from POSSUM and 46 from previous observations\footnote{We confirm that the results presented in this paper do not change significantly even when only the POSSUM sources are considered. This is also true for HI-20 and HI-491.}). We reject 85 sources (light blue colour markers in the bottom right panel of Fig. \ref{fig:main}) that overlap with the strong H$\alpha$ emission region (yellow shades in Fig. \ref{fig:bubbles_zoom}) and use the remaining 289 sources (black markers with error bars) to construct the mean RM profile (red line) and its $1\sigma$ scatter (red shade) enclosing 68\% of the data points in each bin. The blue dashed line and the blue shade are the mean profile and scatter without the $I_{\rm SMC, H\alpha}$ limit for reference. The largest deviation between the profiles with and without the $I_{\rm SMC, H\alpha}$ limit appears between $[0,0.5]\,\theta_{\rm shell, new}$, where the HI-10 shell overlaps with the south-west end of the SMC Bar region. At these angular separations, the $1\sigma$ scatter in the RM distribution is significantly larger without the $I_{\rm SMC, H\alpha}$ limit (blue shade; $\approx80\,\rm rad\,m^{-2}$) compared to the scatter when the $I_{\rm SMC, H\alpha}$ limit is applied (red shade; $\approx20\,\rm rad\,m^{-2}$).

In the RM profile with the $I_{\rm SMC, H\alpha}$ limit, a striking sinusoidal pattern with a period of $3\,\theta_{\rm shell, new}$ is displayed which matches the scale of the entire expanding feature visible in the position--velocity distribution. 
The profile fluctuates around $\approx -10.2\,\rm rad\,m^{-2}$. The RM values at the crest and trough of the mean profile are $\approx 12.8\,\rm rad\,m^{-2}$ and $\approx -34.6\,\rm rad\,m^{-2}$ at $-0.75\,\theta_{\rm shell, new}$ and $+0.75\,\theta_{\rm shell, new}$, respectively. This overall negative RM, even after the Milky Way foreground subtraction, indicates that the large-scale global magnetic field near the HI-10 shell is pointing away from the observer. We measure the RM gradient ($\Delta RM$) at random locations across the entire SMC region and confirm that such a strong gradient ($\Delta RM \approx 50\,\rm rad\,m^{-2}$) over the given angular scale of $\approx1.7^{\circ}$ only occurs around the HI-10 shell and some parts of the SMC Bar region. A typical magnitude of the RM gradient at these random locations is $\lsim 15 \,\rm rad\,m^{-2}$ at this angular scale.

The standard deviation of the residual RM after subtracting the mean profile is $\approx 17.7\,\rm rad\,m^{-2}$. This is significantly larger than the typical error of POSSUM RM measurements ($\approx 1.7\,\rm rad\,m^{-2}$) as well as the RM dispersion of extragalactic origin ($6$ -- $7\,\rm rad\,m^{-2}$) estimated by \citet{Schnitzeler_2010} and \citet{Oppermann_2015}. In Section \ref{sec:model}, we show that at least part of the total excess in the RM variation can come from the turbulent magnetic field in the ISM around the expanding shell. Note that the random component of the magnetic field in the SMC is generally obtained to be quite strong: $\sim 5\,\rm \mu G$ in \citet{Livingston_2022} and $\sim 14\,\rm \mu G$ in \citet{Seta_2023}.
%As reported by \citet{Livingston_2022}, the random component of the magnetic field ($\sim 5\,\rm \mu G$) is strong in the ISM of the SMC in general.

As can be seen in the right panel of Fig. \ref{fig:bubbles_zoom}, the northern edge of the HI-10 overlaps with the HI-20 shell in the (RA, Dec) space (see the green circles with solid line boundary). We also confirm that the two structures intersect at $\approx 140\,\rm km\,s^{-1}$ in the velocity space, indicating that the two shells can be potentially interacting. 
In order to estimate the contribution of the RM measurements that overlap with the HI-20 shell to the results presented above, we show the data points located within the $1\,\theta_{\rm shell, new}$ of HI-20 with square markers in green in the bottom right panel of Fig. \ref{fig:main}. 
We confirm that removing these data points from constructing the RM profile of HI-10 marginally increases the mean profile and reduces the $1\sigma$ scatter between $[-1, 0]\,\theta_{\rm shell, new}$.

\subsection{HI-20 shell}\label{sec:HI-20}

The HI-20 shell is located northwest of the SMC Bar region (see Figs. \ref{fig:shell_sample} and \ref{fig:bubbles_zoom}). GASKAP-HI observations in this area show that it is a site of high-velocity outflows lifting gas into the galaxy-halo interface (\citealt{McClure-Griffiths_2018}). Previously, \citet{Livingston_2022} suggested a possible connection between the outflows and the large RM in this region.

The boundary of the HI-20 shell is characterised by the SMC Bar and the structures on the northeast and southwest sides of the shell surrounding the hole in the HI distribution (the area inside the green dashed circle in the right panel of Fig. \ref{fig:bubbles_zoom}). The HI emission is weaker on the northwest side of the bubble, suggesting that it may be a broken-out galactic chimney.
The southwest wall of the HI-20 is potentially interacting with the HI-10 shell as noted in the previous section.

The axis of the maximum RM gradient across the HI-20 shell is almost perpendicular to the SMC Bar region (see the white dashed line in the right panel of Fig. \ref{fig:bubbles_zoom}). 
Among 169 RM measurements within $2 \theta_{\rm shell, new}$ of the shell centre (144 from POSSUM and 24 from previous studies), 48 sources overlap with the strong H$\alpha$ emission region. 
These sources are distributed throughout the angular separation range presented in Fig. \ref{fig:main} (top right panel) and have a large scatter in the RM distribution, especially around the angular separation $\approx -0.75\,\theta_{\rm shell, new}$ ($\approx74\,\rm rad\,m^{-2}$; see the blue dashed line and the shade).

The mean RM profile with the $I_{\rm SMC, H\alpha}$ limit (red line) is the lowest ($\approx -13.0^{+14.7}_{-16.7}\,\rm rad\,m^{-2}$) near the shell centre, between the angular separation $[0,0.5]\, \theta_{\rm shell, new}$. The profile rises as going outwards from the centre up to $\pm \theta_{\rm shell, new}$. 
We will show in Section \ref{sec:model} that the decrease in the RM near the centre of a shell is typical of expanding magnetised bubbles projected onto the RM grid, while the nearly constant RM profile at larger separations could be due to the effect of other nearby objects, such as the HI-10 shell. 
Indeed, we confirm that most RM measurements with $\rm RM-RM_{\rm MW}>0$ beyond $1\,\theta_{\rm shell, new}$ are within the angular radius of the HI-10 shell (green square symbol).

\subsection{HI-491 shell}

The HI-491 shell is in the SMC Wing region, which connects the SMC to the Magellanic Bridge. 
The HI peak intensity distribution in Fig. \ref{fig:bubbles_zoom} shows long filamentary structures in HI that extend towards the southeast of the SMC. 
There are multiple velocity components presented in the position--velocity diagram ($120-140, 150-160$, and $160-185\,\rm km\,s^{-1}$) corresponding to each filament. 
It has been suggested from the HI kinematics and multi-wavelength observations that there are multiple superbubbles colliding in this region, triggering the formation of young stars at the intersections (\citealt{Nigra_2008, Fukui_2020}). 

It is clear from the visual inspection of the zoom-in panel of Fig. \ref{fig:bubbles_zoom} that a group of positive $\rm RM-RM_{\rm MW}$ measurements at RA (J2000) = $1^{\rm h}26^{\rm m}$ and Dec (J2000) $= -75^{\circ}25'$ overlap with the underlying filamentary structure in HI emission.

There are 247 RM measurements within $2 \theta_{\rm shell, new}$ of the shell centre (139 from POSSUM and 25 from previous studies), and 87 sources overlap with the high H$\alpha$ region in the north of the shell. 
The high H$\alpha$ region is located on the northern side of HI-491 (angular separation $<0$), i.e., towards the direction of the Magellanic Stream, and sources here show a large scatter in the RM distribution ($\approx92\,\rm rad\,m^{-2}$; see the blue dashed line and the shade in the bottom left panel of Fig. \ref{fig:main}).

With the $I_{\rm SMC, H\alpha}$ limit, the RM profile across HI-491 (red line and the shade) is overall enhanced at angular separations $\pm1\theta_{\rm shell, new}$ and gradually decreases going outward from the shell centre toward larger and smaller angular separations. 
The mean RM profile is at its lowest near $\approx-0.25 \theta_{\rm shell, new}$. However, the location of the lowest point shifts away from the centre to $\approx-0.83 \theta_{\rm shell, new}$ when the 1$\sigma$ scatter is considered.
The group of positive RM measurements noted in the visual inspection of the RM grid is located at $0.5 \theta_{\rm shell, new}$. 
%Note that the separations larger than $\gsim \pm 1.25 \theta_{\rm shell, new}$ from this shell are beyond the area covered by the POSSUM pilot field.

\section{Discussion}

\subsection{Expected RM patterns around magnetised bubbles from MHD simulations}\label{sec:model}

\begin{figure*}
    \centering
    \includegraphics[width=\textwidth]{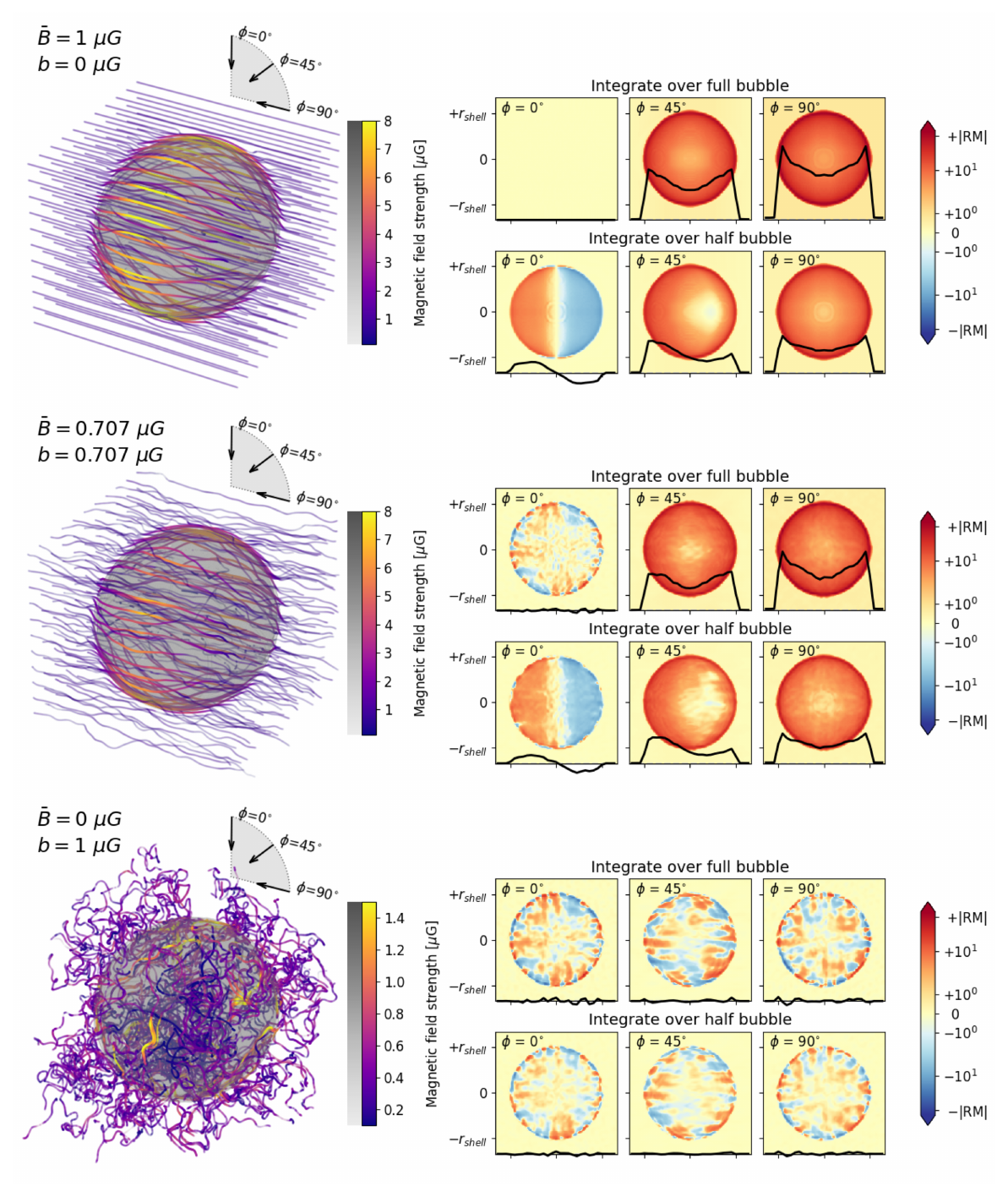}
    \caption{
    Three numerical models of magnetised bubbles with different ambient magnetic field configurations: a uniform mean field ($\boldsymbol{\bar{B}} = 1\,\rm \mu G$, $\boldsymbol{b} = 0\,\rm \mu G$), a composite field ($\boldsymbol{\bar{B}} = 0.707\,\rm \mu G$, $\boldsymbol{b} = 0.707\,\rm \mu G$), and a turbulent random field ($\boldsymbol{\bar{B}} = 0\,\rm \mu G$, $\boldsymbol{b} = 1\,\rm \mu G$). Each model contributes seven panels to the figure. The left panel shows the 3D magnetic field lines in the simulation domain coloured by the magnetic field strength at each point. The isodensity surface of the bubble is shown as a grey sphere. The simulated bubbles do not show elongation along a specific direction within the timeframe covered by the simulations. The right panels are the expected RM distribution around the bubble with different viewing angles (from left to right $\phi = 0^{\circ}$,  $45^{\circ}$, and $90^{\circ}$) and the path length (top: full path through the bubble, bottom: half path through the near side of the bubble). The black line on the bottom of each panel shows the characteristic shape of the mean RM profile along the x-axis and $r_{\rm shell}$ is the radius of the shell. 
    }  
    \label{fig:models}
\end{figure*}

In this section, we use simple numerical models to reproduce various patterns in the RM grid created by magnetised bubbles and discuss their connection to the global magnetic field properties. 
A similar concept has been proposed by \citet{Tahani_2022b, Tahani_2022a,Tahani_2023} where the authors identify that expanding HII bubbles alter the observed plane-of-sky magnetic field around molecular clouds. This suggests that one can infer the original magnetic field configuration at a larger scale before the expansion by studying the magnetic field morphology around the bubbles. Also, \citet{Kothes_2009} demonstrate that expanding supernova remnants can act as magnifying glasses of the ambient magnetic field; the polarized emission from supernova remnants provides information about the large-scale magnetic field configuration of the Galaxy (see also, \citealt{West_2016, Ideguchi_2022}). The major difference between their model and ours is that we intend to simulate the Faraday rotation of polarized radiation coming from sources behind a magnetised bubble of interest. We consider the bubbles to be regions of Faraday rotation without polarized emission. A similar effort has been made by \citet{Purcell_2015} where the authors present a geometric model that matches the RM profile toward the northern part of the Gum Nebula.

We simulate a magnetised numerical bubble by exploding a single supernova in an initially static medium and study how the RM patterns change with the ambient magnetic field configuration variations. For this, we use our \texttt{supernova} module \citep{Buete_2016, Bapna_2019} within a highly modified version of the \texttt{FLASH} code \citep{Fryxell_2000, Dubey_2008} and utilise HLL3R (3-wave approximate) Riemann solver \citep{Waagan_2011}. The numerical simulation consists of a three-dimensional, triply periodic domain of size $200\,{\rm pc}$ sampled on a uniform grid of $144$ grid points along each axis. We initialise our simulations with a uniform gas density of $5\,{\rm cm}^{-3}$, a uniform temperature of $100\,{\rm K}$, and a magnetic field strength of $1\,\mu{\rm G}$. For the initial magnetic field, we consider both a uniform magnetic field, $\boldsymbol{\bar{B}}$, and an isotropic random magnetic field, $\boldsymbol{b}$, which is generated using \texttt{TurbGen} \citep{Federrath_2010, Federrath_2022}. $\boldsymbol{b}$ is constructed to have a power-law magnetic power spectrum of slope $3/2$ \citep[see section 2.3 in][for further details and motivation]{Seta_2020}. We use three representative models with different initial magnetic field configurations: (i) uniform magnetic field ($\boldsymbol{\bar{B}} = 1\,\rm \mu G$, $\boldsymbol{b} = 0\,\rm \mu G$), (ii) random magnetic field ($\boldsymbol{\bar{B}} = 0\,\rm \mu G$, $\boldsymbol{b} = 1\,\rm \mu G$), and (iii) a combination of the uniform and random magnetic field ($\bar{B} = 0.707\,\rm \mu G$, $\boldsymbol{b} = 0.707\,\rm \mu G$). The total magnetic field strength is identical in all three cases by construction and the only difference is in their structure.

After setting up the initial conditions, for each magnetic field configuration, we inject energy of $10^{51}\,{\rm ergs}$ with the Sedov-Taylor solution \citep[][which implies that the $72\%$ of the injected energy goes to the thermal energy and remaining 28\% goes to the kinetic energy]{Sedov_1959} into $16^3$ grid points at the centre of the domain to simulate a supernova explosion and then solve the magnetohydrodynamic equations with heating and cooling functions \citep[see Section 2 in][for details]{Seta_2022} to track the evolution. The explosion expands into the medium and enhances the magnetic field in the shell. The simulations are run for $0.33\,{\rm Myr}$ and we take the final state as representative magnetised bubbles for all three cases. The left panel of Fig. \ref{fig:models} shows the magnetic field lines in 3D, coloured by the field strength at each location. The field line paths are integrated using a fourth-order Runge-Kutta method\footnote{We use {\sc yt project} (\citealt{Turk_2011}) for this analysis.}. 
The path starting points are distributed uniformly in a grid at one side of the box for the uniform and composite field models. For the random field model, we randomly located them within the simulation domain. The iso-density surface of the bubble is shown as a grey sphere surrounded by the field lines.

We use gas density and temperature to determine the thermal electron density \citep[see equation 15 in][]{Hollins_2017} to compute RM from the simulated bubbles. Using these models, we study the effect of varying the viewing angle, the asymmetry between the near and far sides of a bubble, and the coherency of the global ambient magnetic field on the observable RM pattern around the bubbles.

On the right side of Fig. \ref{fig:models}, we present a set of six panels showing the 2D projections of the models to the RM grid.
From left to right, we vary the viewing angle, where $\phi = 0^{\circ}$ is when the sightline and the initial mean field are perpendicular and $\phi = 90^{\circ}$ is when they are parallel (specifically, when the magnetic field points towards the observer). The black line at the bottom of each panel shows the mean RM profile in one dimension, along the x-axis of the figures. 
The upper panels are the RM distributions when integrated over the full path length of the bubble and the bottom panels are the RMs when only the near side of the bubble is considered. This half-bubble integration is designed to test the effect of asymmetry between the near and far sides of the bubbles. Such asymmetry in the bubble's structure can be caused by, for example, a density gradient in the ambient ISM (\citealt{Kompaneets_1960, Basu_1999, Stil_2009}). The half-bubble integration only through the near side corresponds to the extreme case when the magnetic field or the density of the far side is completely null. 

Expanding bubbles sweep up and bend the ambient magnetic fields in all numerical models. 
The behaviour of the magnetic field lines due to the bubble expansion is physically expected to be directly proportional to the expansion velocity, $v_{\rm exp}$ (ambient magnetic fields will be more easily affected by a stronger explosion), and inversely proportional to the Alfven velocity (stronger ambient magnetic field would more strongly resist the effect of expansion), $v_{\rm A} = B/\sqrt{4\upi\rho}$ where $B$ and $\rho$ are the magnetic field strength and density of the medium the bubble is expanding into. While the choice of parameters for the models is based on Milky Way-type conditions, we take the ratio between the two velocities ($v_{\rm exp} / v_{\rm A}$) and use it as a measure of how effective the bending of magnetic fields is as a result of the expansion. For the three models we present in this work, $v_{\rm exp} / v_{\rm A}\approx 25$. The observed expansion velocities of the three bubbles (HI-10, HI-20, and HI-491) identified in Section \ref{sec:shell_catalogue} are $v_{\rm exp} = 22, 12$, and $22\,\rm km\,s^{-1}$, respectively. 
If we assume that these bubbles have the same $v_{\rm exp} / v_{\rm A}$ with the numerical models we present, the corresponding strength of the ambient magnetic field (ordered and random components combined) is $\sim 0.1\,\rm \mu G$.

In the uniform field model, the magnetic field strength on the shell surface is largest along the equatorial plane perpendicular to the initial field. The composite field model follows a similar trend but with less significant enhancement in the magnetic field strength. While the random field model still shows enhanced magnetic fields at random locations of the shell surface, the overall effect is much less prominent than the other models (in Fig. \ref{fig:models}, note the change in the colour scales showing the magnetic field strength in the bottom panel). 
Although not depicted in the figure, the electron density is also higher at the shell surface, contributing to the larger values of RM, as we will show below.

The models show distinctive patterns when projected to the RM grid. We first describe and compare the patterns seen in each field configuration and different viewing angles. Then, we use the lessons from the models to interpret the observed RM profiles across the SMC bubbles.
First, we focus on the uniform field model (top panels in Fig. \ref{fig:models}). Due to the perfect axial symmetry of the uniform field model, the Faraday rotation at the bubble's near and far sides cancels out along the sightlines when $\phi = 0^{\circ}$. In this case, the net RM is zero, and the magnetised bubble does not show up in the RM grid. When only the near half is considered, a clear sign-change in the RM is visible across the bubble. In this case, the axis of the maximum RM gradient, i.e., the axis dividing the positive and negative sides, is perpendicular to the direction of the plane-of-sky magnetic field of the surrounding medium. In other cases where there is a sizable line-of-sight component of the global mean magnetic fields ($\phi = 45^{\circ}$ and $90^{\circ}$), both full and half-length integrations through the bubble create RM profiles that are enhanced along the rim with a rapid drop outside the boundary of the bubble. The location of the minimum |RM| within the projected bubble is the centre of the bubble when the full bubble is considered, while for the half-bubble cases, it tends to shift from the centre to one side as the viewing angle shifts from $\phi = 90^{\circ}$ to lower.

In the composite field model (middle panels), the random field component breaks the perfect symmetry of the mean field. The full bubble-length integration along the $\phi = 0^{\circ}$ viewing angle gives rise to small-scale fluctuations in the RM grid. 
Other choices of $\phi$ show large-scale RM patterns that are largely similar to the uniform model, with added small-scale variations. 

Without the presence of a large-scale mean field, the random field model (bottom panels) does not reveal any large-scale patterns in the RM grid. The imprint of the magnetised bubble appears as enhanced small-scale RM fluctuations in the region. There is no significant change in the RM distribution with varying viewing angles, as expected.

Here, we give the interpretations of the observed RM profiles of the HI-10, 20, and 491 shells from what we have learned from the simple numerical models:
\begin{itemize}
    \item According to our models, the strong RM gradient across the HI-10 shell is reproduced only when (i) the global mean magnetic field in the ambient ISM is coherently aligned along the plane-of-sky at the angular scale of the shell and (ii) the near and far sides of the bubble are not in perfect symmetry; otherwise, Faraday rotation at each side will cancel out. Yet, there is substantial scatter in the residual RM distribution, as discussed in the previous section. This indicates that the magnetic field around the HI-10 shell also has a random component, like the composite field model shown in this section. A similar RM pattern (a sign-change at the scale of a bubble) is reported by \citet{Harvey-Smith_2010} and \citet{Heald_2012}. For example, in the latter, the authors find a strong RM gradient at one of the HI holes in a nearby face-on galaxy NGC 6946. 
    
    \item The HI-20 shell shows a RM profile that increases going towards the edge of the shell from the centre up to approximately $\pm 1\theta_{\rm shell, new}$. This is a typical pattern in our models when the ambient mean magnetic field has a sufficient line-of-sight component at scales comparable to/larger than the bubble and/or the far and near sides of the bubble are largely symmetric. The difference we find between the numerical models and the observed RM profile of the HI-20 shell is that the rapid drop in the mean RM outside the shell boundary is not visible in the observed profile. This is potentially because the HI-20 shell is located in the vicinity of the SMC Bar region, where active star formation takes place. It is very likely the area at large angular separations from the HI-20 shell centre is contaminated by other magneto-ionised superbubbles, such as the HI-10 shell, that can also increase the overall RM distribution.

    \item The HI-491 shell shows a RM profile that is enhanced at the shell edges and drops rapidly outside the shell, indicating that the substantial portion of the mean magnetic field is coherently ordered along the line of sight. This aligns with previous studies by \citet{Kaczmarek_2017} and \citet{Livingston_2022} that reported the presence of a large-scale mean magnetic field in the SMC Wing region and the Magellanic Bridge that is overall pointing away from the observer. The formation of this coherent large-scale field is potentially an outcome of tidal interactions between the Magellanic Clouds, resulting in the formation of the Magellanic Bridge (\citealt{Nidever_2013}). 
    The lowest point of the RM profile being slightly offset from the bubble's centre may suggest that the near and far sides of the bubble may not be perfectly symmetric. Yet, the significance of the offset needs to be tested by future studies with higher RM source densities in this area. Even with POSSUM's high RM source density, we could not use bins smaller than $0.5\,\theta_{\rm shell, new}$ while ensuring enough RM measurements in each bin. 
\end{itemize}

There are two possible mechanisms a foreground bubble can alter the observable RM distribution of background sources. First, the polarized emission from the bubble itself may interfere with the emission from the background sources and affect our ability to reliably extract RM. The diffuse subtraction step explained in Section \ref{sec:method-possum} removes most of the foreground emission from the shells and, therefore negates this effect. Second, turbulent RM fluctuations smaller than the beam will cause beam depolarization of background sources. This could reduce the RM source density but should not change the observed RM as long as the mean RM fluctuation in this turbulent medium is zero, which is normally assumed for a path length greater than $100\,\rm pc$. Therefore, the shape of the model RM profiles would not change as a result of beam depolarization.

While we find a promising resemblance between the RM patterns around the observed HI bubbles and the simple numerical models, the question of whether the RM patterns are indeed a universal indicator of the ambient magnetic field configuration should be tested further using other direct and indirect tracers of magnetic fields. For example, \citet{Chen_2022} suggest that the position angle of an elongated bubble reveals the global plane-of-sky magnetic field orientation. Although the \citetalias{Staveley-Smith_1997} catalogue does not provide information for the elongation of the HI shells, the HI-10 and HI-491 shells have their counterpart in the \citetalias{Stanimirovic_1999} catalogue where the ellipticity measurements are available: sgs-37A and sgs-494A (see the pink ellipses in Fig. \ref{fig:shell_sample}). The ellipticities of sgs-37A and sgs-494A are 0.45 and 0.56, respectively.

The position angle of sgs-37A is $160^{\circ}$, measured from north to east on the image plane. Comparing the position angle with the axis of the null RM gradient of the HI-10 shell, we find that the minor axis of sgs-37A and the RM gradient axis are offset by $\sim 39^{\circ}$. As can be seen from the uniform model, the axis of the null RM gradient is perpendicular to the direction of the magnetic plane-of-sky magnetic field when $\phi\approx0^{\circ}$. The misalignment of the two axes in the observed bubble can be due to (i) the uncertainty in defining the position angle and the RM axis and (ii) the presence of a random component in the global magnetic field. For reference, in the composite field model, where we introduce the random field that breaks the perfect symmetry of the uniform field model, the two axes are offset by $\approx3^{\circ}$.

Sgs-494A, which encompasses the HI-491 shell, is elongated along the direction from the SMC towards the Magellanic Bridge. The position angle of this SGS is $150^{\circ}$. The angle between the minor axis of the sgs-494A and the axis of the maximum RM gradient is $\sim 25^{\circ}$. 
From the shape of the RM profile that rises towards the bubble's edge, we predict the presence of a sizable line-of-sight component in the global mean magnetic field. 
If the plane-of-sky component is also present, our models predict that the minimum point of the RM profile would be offset from the centre. 
While we conclude that higher density RM grids with, e.g., the Square Kilometre Array (SKA) are required to confirm this, the relatively small angle between the bubble's minor axis and the axis of the maximum RM gradient also supports the existence of a magnetic field component aligned with the plane-of-sky.

We postpone investigations of direct observational tracers of plane-of-sky magnetic fields, such as starlight and dust polarization, to future studies. This will provide a comprehensive understanding of the local magnetic field structure around the bubbles as well as the global magnetic field in the SMC's ISM in general.

\subsection{Magnetic field strength in the HI shells}\label{sec:B_strength}

\begin{figure*}
    \centering
    \includegraphics[width=0.85\textwidth]{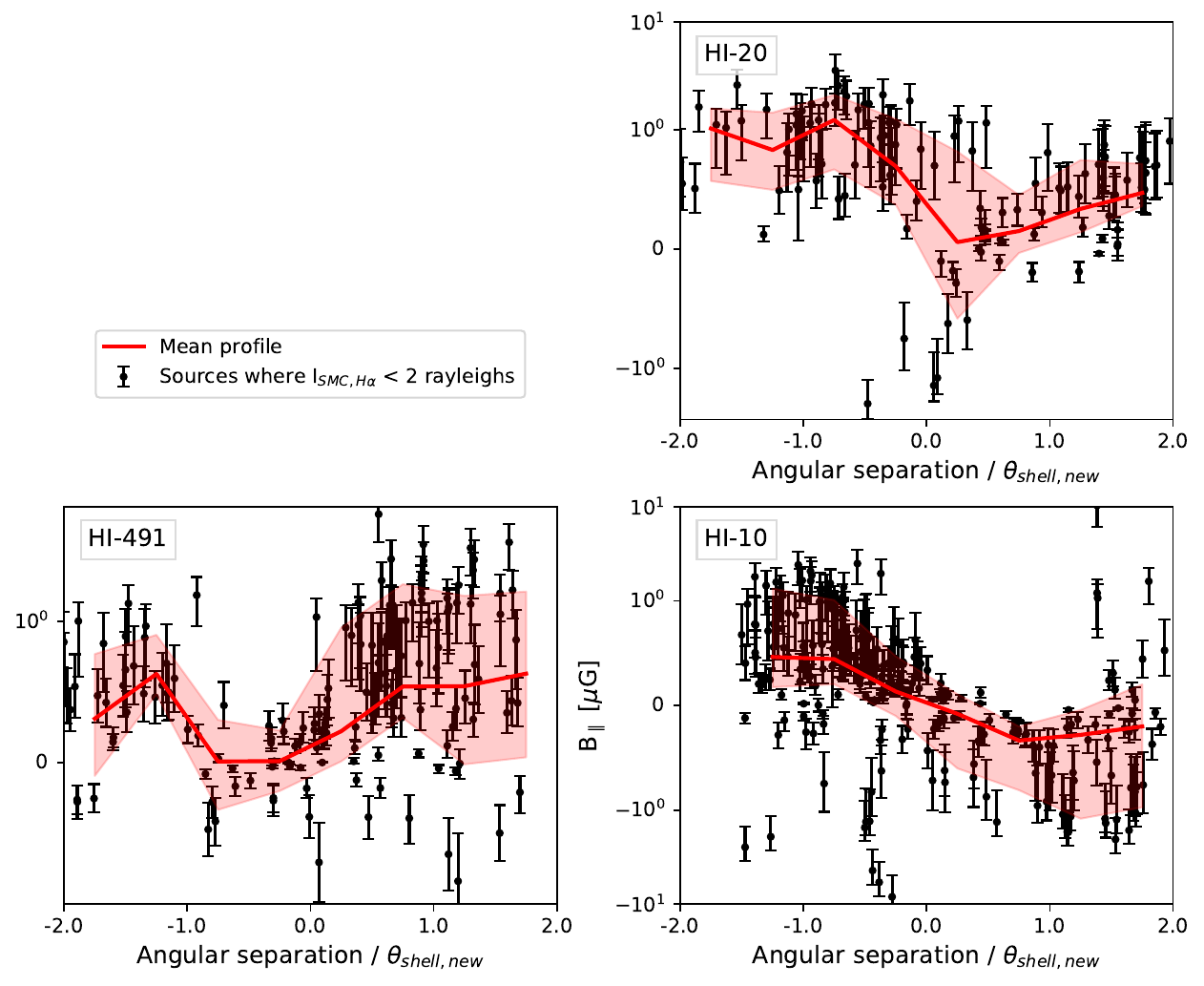}
    \caption{
    The $B_{\rm \parallel}$ profiles across the HI-10, HI-20, and HI-491 shells. The y-axis is linear between $\pm1\,\rm \mu G$ and in a log-scale otherwise. We exclusively use RM measurements within the GASKAP-HI SMC field and in the low H$\alpha$ emission region to calculate $B_{\rm \parallel}$ (black symbols with error bar). The red line and shade are the mean and the $1\sigma$ scatter of the profile binned in $0.5\,\theta_{\rm shell, new}$ size between $[-2, 2]\,\theta_{\rm shell, new}$ range. Overall, the HI shells enhance the line-of-sight magnetic fields to the order of $\sim 1\rm\mu G$.
    }
    \label{fig:b_los}
\end{figure*}

In this Section, we estimate line-of-sight magnetic field strength in the HI shells from the observed RM measurements. First, we further refine the RM produced purely by the bubbles. Although we have already taken care of the Milky Way foreground effect as described in Section \ref{sec:foreground}, the RM measurements need to be shifted by a certain amount before being converted to magnetic field strength. This is because the $\rm RM-RM_{\rm MW}$ values still incorporate the RM produced by the SMC, as well as the possibility of imperfect Milky Way foreground subtraction. Therefore, we define the reference RM ($\rm RM_{\rm ref}$) of each bubble and use $\rm RM-RM_{\rm MW}-RM_{\rm ref}$ to calculate the magnetic field strength. Ideally, we should be able to obtain $\rm RM_{\rm ref}$ by taking RM sources away from a bubble of interest but still located within the SMC (e.g., \citealt{Tahani_2018}). However, the ISM structure of the SMC is chaotic, and there are no ``clean'' off-bubble regions (see Fig. \ref{fig:shell_sample}) from which we can draw the reference sources. Thus, we instead make assumptions based on the numerical models presented in the previous section. 

In model RM grids where the sign of the magnetic field changes across a bubble (i.e., uniform/composite field, $\phi = 0^{\circ}$), the RM profile fluctuates around zero. We assume that this is the case for the HI-10 shell. As the mean RM profile of the HI-10 shell fluctuates around $-10.2\,\rm rad\,m^{-2}$, we take this value as $\rm RM_{\rm ref}$ in this region\footnote{We take $\rm RM_{\rm ref}=-10.2\,\rm rad\,m^{-2}$ as this value equalises the area enclosed by the positive and negative sides of the $\rm RM-RM_{\rm MW}-RM_{\rm ref}$ profile.}. On the other hand, in model RM grids where the RM profile peaks at the bubble boundary, the true $\rm RM_{\rm ref}$ can only be recovered at large angular separations outside the bubbles, where the RM profile drops rapidly. However, we cannot recover $\rm RM_{\rm ref}$ from outside the bubble for the reason above. As an alternative, we take the minimum point of the RM profile near the centre of the shells as $\rm RM_{\rm ref}$ ($-13.0\,\rm rad\,m^{-2}$ for HI-20 and $-21.1\,\rm rad\,m^{-2}$ for HI-491). This way, we are measuring the relative enhancement in the line-of-sight magnetic field strength at the bubble's edges with respect to the centre.

Information about the electron density distribution along sightlines is required to estimate the strength of the line-of-sight magnetic field from the RM measurements. Based on the assumption that the electron density and magnetic field strength are uncorrelated \citep{Seta_2021}, equation \ref{eq:RM} can be expressed as follows.
\begin{equation}\label{eq:b_los}
    \left(\frac{B_{\rm \parallel}}{\rm \mu G}\right) \approx \frac{1}{0.812}\left(\frac{\rm RM}{\rm rad\,m^{-2}}\right)\left(\frac{\rm DM}{\rm pc\,cm^{-3}}\right)^{-1},
\end{equation}
where ${\rm DM} \equiv \int n_{\rm e} {\rm d}l$. 
This is a reasonable assumption for a typical diffuse ISM. However, the compressive motion of expanding superbubbles enhances both the electron density and magnetic field. 
Such positive correction between $n_{\rm e}$ and $B_{\rm \parallel}$ could overestimate the mean field strength by a factor of $2-3$ (\citealt{Beck_2003, Seta_2021}).

\citet{Livingston_2022} compare various models of the electron density distribution of the SMC and conclude that the model directly relating the HI column density and the pulsar dispersion measure (DM) provides the most reliable magnetic field estimates (see their Section 4). 
This paper follows their decision, and we only briefly summarise the model here. 
Assuming that the electron density is directly characterised by the neutral hydrogen density and its ionisation fraction ($X_{\rm e}$), the DM is related to the HI column density as
\begin{equation}\label{eq:dm}
    \left(\frac{{\rm DM}_{\rm pulsar}}{\rm pc \, cm^{-3}}\right) = 3.24\times 10^{-19}\,X_{\rm e} \left(\frac{N_{\rm HI}}{\rm cm^{-2}}\right).  
\end{equation} 
The empirical relation between the pulsar DM measurements and the corresponding HI column density in the SMC region is 
\begin{equation}\label{eq:dm_smc}
    \left( \frac{N_{\rm HI, SMC}}{\rm 10^{20}\, cm^{-2}} \right) = (0.15\pm0.06)\left(\frac{{\rm DM}_{\rm pulsar, SMC}}{\rm pc \, cm^{-3}}\right).
\end{equation}
Equating equations \ref{eq:dm} and \ref{eq:dm_smc} gives $X_{\rm e} = 21^{+16}_{-6}\%$. 
The model generalises the relation given in equation \ref{eq:dm_smc} at sightlines without ${\rm DM}_{\rm pulsar, SMC}$ measurements. 
At the location of each RM source, we take the HI column density from the GASKAP-HI SMC data and calculate the expected DM. 
RM measurements outside the GASKAP SMC field are neglected for further analysis. 
Then, we calculate $B_{\rm \parallel}$ based on equation \ref{eq:b_los}. 
Note that we assume the entire HI column density in GASKAP-HI in these sightlines is associated with the bubbles.
If other unrelated HI structures exist in these directions, the true electron densities of the bubbles would be lower than our estimation.
That is, we are potentially underestimating the magnetic field strength associated with the bubbles. 
Another point to note is that, physically, the volume within the HI shells is expected to be filled with hot, fully ionised gas. The true ionised fraction through the HI voids of the shells could be larger than the ionised fraction of 21\% from the pulsar DM measurements. In this regard, we are potentially overestimating $B_{\rm \parallel}$ for sightlines closer to the shell centre. 

Fig. \ref{fig:b_los} shows $B_{\rm \parallel}$ profiles across each bubble. The y-axes of the panels have a scale which is linear between $\pm1\,\rm \mu G$, log otherwise. 
The error of each $B_{\rm \parallel}$ measurement is a combination of the RM error and the error of the slope of the $N_{\rm HI, SMC}$--$\rm DM_{\rm pulsar, SMC}$ relation. 
The line-of-sight magnetic field strength in the HI-10 shell (bottom right panel) monotonically increases, going towards negative angular separation from positive angular separation. The scatter is larger near the outer boundary of the bubble ($\pm 2 \theta_{\rm shell, new}$). The mean $|B_{\rm \parallel}|$ in this range is $\sim 0.5\,\rm \mu G$, but individual sightlines can go up to the order of $1\,\rm \mu G$ in some cases. In both the HI-20 and HI-491 shells (top right and bottom left panels, respectively), we find $\sim 1\,\rm \mu G$ enhancement in the line-of-sight magnetic field strength with respect to the central regions of the bubbles.

In summary, we find $|B_{\rm \parallel}|\sim 1\,\rm \mu G$ enhancement at the edges of all three HI shells, while the exact shape of the $B_{\rm \parallel}$ profile can vary.
Based on the simple numerical models presented in the previous subsection, we suggest that the enhancement in $|B_{\rm \parallel}|$ is because the magnetic field is getting stronger and more ordered in the compressed region, as the magnetic field lines are stretched and compressed \citep{Setab_2020, Seta_2021b} due to the expansion of the shell while the shape of the profile depends on the viewing angle and the surrounding environment.

\section{Summary}\label{sec:summary}

In this paper, we report the first detection of magnetic fields associated with HI superbubbles in the SMC: HI-10, HI-20, and HI-491. 
Using the POSSUM pilot survey, we obtain a high-density RM grid ($\approx 25\,\rm deg^{-2}$) towards the SMC. 
The magnetised bubbles produce large-scale patterns in the RM distribution at angular scales comparable to the extent of the bubbles in HI emission. 
We show that our numerical models of expanding bubbles in a magnetised medium reproduce the overall shape of the observed RM profile across the HI bubbles.
The models predict that the observable RM profile depends on (i) the presence of coherent and random magnetic field components, (ii) the asymmetry in the bubble’s magnetic field structure at the near and far sides from the observer, and (iii) the viewing angle.

\begin{enumerate}
    \item The large-scale pattern in the RM profile comparable to the size of a bubble is only visible when the bubble is expanding within a medium with coherently ordered magnetic fields at scales larger than the bubble. When the ambient magnetic field is completely random, the imprint of magnetised bubbles appears as increased scatter in the RM distribution within the region. Our observational finding that all three HI bubbles investigated in this paper show large-scale patterns in the RM profile indicates that there is a large-scale coherent magnetic field in the SMC at the vicinity of the bubbles. However, the large scatter in the RM distribution with respect to the mean profile suggests the presence of a substantial random component as well.
    \item The sign change in the RM profile with respect to the centre of a bubble can only be found when the near and far sides of the bubble are asymmetric. When the bubble is in perfect symmetry, the RM profile is always symmetric around the centre of the bubble. The sign-changing RM pattern across the HI-10 shell indicates the asymmetry of the bubble's structure at the near and far sides.
    \item The shape of RM profiles changes depending on the viewing angle of a bubble with respect to the ambient large-scale magnetic field. When the ambient field is overall parallel to the line-of-sight, the RM profile is symmetric with a decline near the centre. As the angle between the sightline and the ambient large-scale field increases, the shape of the RM profile becomes more asymmetric, given the asymmetry of the bubble's near and far sides. When the viewing angle is perpendicular to the ambient field, we find a sign change in the RM profile with respect to the centre. Based on this, we suggest that the large-scale field around the HI-10 shell is overall aligned along the plane of the sky, while the field configuration around the HI-20 and HI-491 shells incorporates substantial line-of-sight components.

\end{enumerate}

In the three HI shells analysed in this paper, we find that the magnetic field strength is enhanced at the shell edges by $\sim 1\,\rm \mu G$ with respect to the shell centre as a result of the compression of expanding bubbles. This is about an order of magnitude stronger than the ambient magnetic field strength ($\sim 0.1\,\rm \mu G$, ordered and random components combined) estimated based on the expansion velocity of the shells. The degree of enhancement is in agreement with the compression factor expected from previous models of an expanding shell (\citealt{vanderLaan_1962}).

\section*{Acknowledgements}

We thank the referee for their helpful report which significantly improved the paper.
This scientific work uses data obtained from Inyarrimanha Ilgari Bundara, the CSIRO Murchison Radio-astronomy Observatory. We acknowledge the Wajarri Yamaji People as the Traditional Owners and native title holders of the Observatory site. CSIRO’s ASKAP radio telescope is part of the Australia Telescope National Facility (https://ror.org/05qajvd42). Operation of ASKAP is funded by the Australian Government with support from the National Collaborative Research Infrastructure Strategy. ASKAP uses the resources of the Pawsey Supercomputing Research Centre. Establishment of ASKAP, Inyarrimanha Ilgari Bundara, the CSIRO Murchison Radio-astronomy Observatory and the Pawsey Supercomputing Research Centre are initiatives of the Australian Government, with support from the Government of Western Australia and the Science and Industry Endowment Fund. The POSSUM project (https://possum-survey.org) has been made possible through funding from the Australian Research Council, the Natural Sciences and Engineering Research Council of Canada, the Canada Research Chairs Program, and the Canada Foundation for Innovation.

S.L.J acknowledges support from the UK Research and Innovation (UKRI) Frontiers Research Grant [EP/X026639/1]. 
This work was partially funded by the Australian Government through an Australian Research Council Australian Laureate Fellowship (project number FL210100039) to N.Mc-G and through the Discovery Projects funding scheme (project DP220101558). 
J.D.L is part of the M2FINDERS project. The M2FINDERS project has received funding from the European Research Council (ERC) under the European Union’s Horizon 2020 research and innovation programme (grant agreement No 101018682). 
B.M.G. acknowledges the support of the Natural Sciences and Engineering Research Council of Canada (NSERC) through grant RGPIN-2022-03163 and of the Canada Research Chairs program. 
M.T. is supported by the Banting Fellowship (Natural Sciences and Engineering Research Council Canada) hosted at Stanford University and the Kavli Institute for Particle Astrophysics and Cosmology (KIPAC) Fellowship.
C.F. acknowledges funding provided by the Australian Research Council (Discovery Project DP230102280), and the Australia-Germany Joint Research Cooperation Scheme (UA-DAAD).
S.P.O. acknowledges support from the Comunidad de Madrid Atracción de Talento program via grant 2022-T1/TIC-23797.

We further acknowledge high-performance computing resources provided by the Leibniz Rechenzentrum and the Gauss Centre for Supercomputing (grants~pr32lo, pr48pi and GCS Large-scale project~10391), the Australian National Computational Infrastructure (grant~ek9) and the Pawsey Supercomputing Centre (project~pawsey0810) in the framework of the National Computational Merit Allocation Scheme and the ANU Merit Allocation Scheme.

Our analysis was performed using the Python programming language (Python Software Foundation, https://www.python.org). The following packages were used throughout the analysis: numpy (\citealt{Harris_2020}),  SciPy (\citealt{Virtanen_2020}), matplotlib (\citealt{Hunter_2007}), and yt project \citep{Turk_2011}.

\section*{Data availability}
This paper uses raw data products that are publicly available on the CSIRO ASKAP Science Data Archive (CASDA; \citealp{Chapman2017,Huynh2020}).

%%%%%%%%%%%%%%%%%%%%%%%%%%%%%%%%%%%%%%%%%%%%%%%%%%



\bibliographystyle{mnras}
\bibliography{main} % if your bibtex file is called example.bib

%\appendix
%\section{Selection criterion for HI shells}\label{sec:selection}

% Don't change these lines
%\bsp	% typesetting comment
%\label{lastpage}
\end{document}